\documentclass[11pt,aps,pra]{revtex4-1}
\usepackage{graphicx,bm}

\begin{document}

\title{Inelastic low-energy collisions of electrons with HeH$^+$: rovibrational excitation
and dissociative recombination}

\author{Roman \v{C}ur\'{i}k} \email{roman.curik@jh-inst.cas.cz}
\affiliation{J.~Heyrovsk\'{y} Institute of Physical Chemistry, Academy of Sciences
of the Czech Republic,~v.v.i., Dolej\v{s}kova~3, 182~23 Prague~8, Czech Republic}
\author{Chris~H.~Greene}
\affiliation{Department of Physics and Astronomy, Purdue University, West Lafayette,
Indiana 47907, USA}

\date{\today}

\begin{abstract}
Inelastic low-energy (0--1 eV) collisions of electrons with HeH$^+$
cations are treated theoretically, with a focus on the rovibrational excitation and dissociative recombination (DR)
channels. In an application of {\it ab initio} multichannel quantum defect theory (MQDT),  the description of both processes is based on the Born-Oppenheimer quantum defects.
The quantum defects were determined using the R-matrix approach in two different frames
of reference: the center-of-charge and the center-of-mass frames. The results obtained in the two reference systems, after implementing the Fano-Jungen style rovibrational frame-transformation technique, shows differences in the rate of convergence
for these two different frames of
reference. We find good agreement with the available theoretically predicted rotationally inelastic thermal rate coefficients. Our computed DR rate also agrees well with available experimental results. Moreover, several computational experiments
shed light on the role of rotational and vibrational excitations in the indirect
DR mechanism that governs the low energy HeH$^+$ dissociation process. While the rotational excitation is several
orders of magnitude more probable process at the studied collision energies, the closed-channel
resonances described by the high-$n$, rotationally excited neutral molecules of HeH 
contribute very little to the dissociation probability. But the situation is very different for
resonances defined by the high-$n$, vibrationally excited HeH molecules, which are found
to dissociate with approximately 90\% probability.
\end{abstract}

\maketitle

\section{\label{sec-Intro}Introduction}

Low-energy collisions of electrons with HeH$^+$ form one of the many ingredients
necessary to understand the chemistry of the early universe \cite{Roberge_Dalgarno_HeH_1982}.
Furthermore, the relative simplicity of the system helps to gauge new experimental setups
while also serving as a benchmark for different theoretical models. For these reasons,
the low-energy collision process between electrons and the HeH$^+$ cations has attracted experimental interest
\cite{Yousif_Mitchell_1989,Yousif_Mitchell_1994,Sundstrom_Larsson_1994,
Tanabe_Takagi_HeH_1994,Mowat_Larsson_HeH_1995,Stromholm_Larsson_HeH_1996,
Semaniak_Zajfman_HeH_1996,Tanabe_Takagi_HeH_1998}
and theoretical studies 
\cite{Yousif_Mitchell_1994,Guberman_HeH_1994,Sarpal_Morgan_HeH_1994,
Takagi_HeH_2004,Haxton_Greene_HeH_2009,Takagi_Tashiro_2015,Hamilton_Tennyson_HeH_2016}
over the past three decades.

The dissociative recombination (DR) of HeH$^+$ 
\begin{equation}
\mathrm{HeH}^+ + \mathrm{e}^- \rightarrow \mathrm{He} + \mathrm{H}(n)\;,
\end{equation}
belongs to a class of processes for which
no curve crossing of the target cation and the dissociating neutral exist. The resulting
indirect mechanism often led to underestimates of the DR 
rates in early theoretical studies \cite{Roberge_Dalgarno_HeH_1982}. It was established later
\cite{Guberman_HeH_1994,Curik_Greene_PRL_2007,Haxton_Greene_HeH_2009,Curik_Gianturco_2013} that the
light diatomic cations do not always require the presence of a curve crossing to produce a high DR cross section.
Multi-channel quantum defect theory (MQDT) helped to understand that the indirect DR
is driven by transitions to auto-ionizing Rydberg states attached to a rovibrationally
excited ionization threshold. The nuclear dynamics is then driven by non-adiabatic coupling
among these Rydberg states. The cross section exhibits series of sharp peaks that accumulate
at the rovibrational thresholds. The corresponding auto-ionizing Rydberg states are also referred to in the literature as
rovibrational Feshbach resonances.

The second inelastic process that is energetically allowed at very low energies is rovibrational
excitation by electron impact
\begin{equation}
\mathrm{HeH}^+(\nu' j') + \mathrm{e}^- \rightarrow \mathrm{HeH}^+(\nu j) + \mathrm{e}^-\;.
\end{equation}
Measurement of the rovibrationally inelastic cross sections is a challenging
experimental task, owing largely to the required high electron energy resolution. On the other hand, these cross sections constitute a key
ingredient of models designed to determine the initial state populations
of the cations in storage rings 
\cite{CSR_review_2016,Zajfman_Ullrich_conf_2005}. While cooling of the target
cations in the black body radiation environment should eventually lead to a stable Boltzmann distribution, the
additional inelastic collisions with the electrons can modify and establish quite
different rovibrational populations of the cations. 
Despite its importance, there appears to only be a single previous study that reports 
the rotationally inelastic cross sections for e$^-$ + HeH$^+$,
namely \cite{Hamilton_Tennyson_HeH_2016}.

Several theoretical methods have been applied to treat the non-adiabatic recombination nuclear dynamics
in electron collisions with the HeH$^+$ cation, with various levels of success.
Guberman in his pioneering work \cite{Guberman_HeH_1994} adopts a traditional description 
of the non-adiabatic nuclear dynamics. He identifies the discrete-continuum coupling
via avoided crossing of the C state (open channel) of the neutral HeH and its D state 
that lies in the closed-channel space. The coupling between the target cation and the
Rydberg series of the neutral is described using quantum defect theory (QDT). 
Rotational degrees of freedom were not included in the Guberman study and their
importance for the dissociative and inelastic processes remains to be investigated.

The theoretical methodology used by Takagi 2004 \cite{Takagi_HeH_2004,Takagi_Tashiro_2015}
is closer in spirit to the present approach. Takagi constructed his Born-Oppenheimer
wave function at short electronic distances from the fixed-nuclei quantum defects. 
Following a rovibrational frame transformation (FT), the 
wave function was expressed in terms of the asymptotic (or channel) quantum numbers. Then the
asymptotic boundary conditions were imposed by the usual MQDT closed channel elimination procedure.
 The main difference from the present study lies in the 
description of the nuclear continuum. Takagi used real-valued box states that seem to be
connected to the asymptotic dissociative states by a simultaneous renormalization
of the nuclear and electronic parts of the total wave function.
In the second, more recent study, Takagi and
Tashiro \cite{Takagi_Tashiro_2015} carried out two different studies
that utilized different coordinate system centers. They reported that the DR cross 
sections computed in the center-of-mass system (CMS) are about two orders of magnitude
smaller than the cross section computed in the center-of-charge system (CCS). Since the
choice of the coordinate system is important for justifying the validity of various approximations
used in the preset MQDT approach, we also address this issue in the present study.

The computational procedure used by Haxton and Greene 
\cite{Haxton_Greene_HeH_2009} is quite similar to the present study.  The DR results presented
here may be considered as a refinement of those published by Haxton and Greene, as most of
the parameters defining the calculations were obtained in that study, and extended to some degree in the present treatment. In particular, 
the partial
waves employed here are limited by $l_{max}$ = 5, whereas $l_{max}$ = 3 was used in
the previous calculations. Moreover, the nuclear dynamics is described here on an
interval of interatomic distances from 0.2 to 10 a.u., while \cite{Haxton_Greene_HeH_2009} used the interval from 0.8 to 5.0 a.u. Finally, more
notable difference can be found in the choice of vibrational functions. 
Ref.\cite{Haxton_Greene_HeH_2009} employed C-normalized complex 
functions resulting from the exterior complex scaling method
\cite{Moiseyev_CS_1998} that satisfy conventional C-orthogonality relations, while
our model uses Siegert pseudo-states as will be described below.

The present study is motivated by ongoing experimental needs (specifically, a need for the inelastic collision
rates) and also by a goal to understand and resolve some computational discrepancies among the different theoretical approaches
mentioned above. In order to address these points, we have formulated the following goals.
\begin{itemize}
\item[a.] Determination of accurate CMS and CCS quantum defects of HeH$^+$ as a function
          of internuclear distance and explanation of the differences between the CMS and
          CCS approaches
\item[b.] Computation of rovibrationally inelastic rates for the lowest rovibrational
          transitions
\item[c.] Test the credibility of our model by computing the DR rates and comparing them with the available experimental data
\item[d.] Analysis of the origin of the dominant low-energy DR peaks and discussion of the
          DR mechanism for electron collisions with the HeH$^+$ cation
\end{itemize}

\section{\label{sec-QDs} Quantum defects}

\begin{figure}[tbh]
\begin{center}
\includegraphics[width=0.6\linewidth]{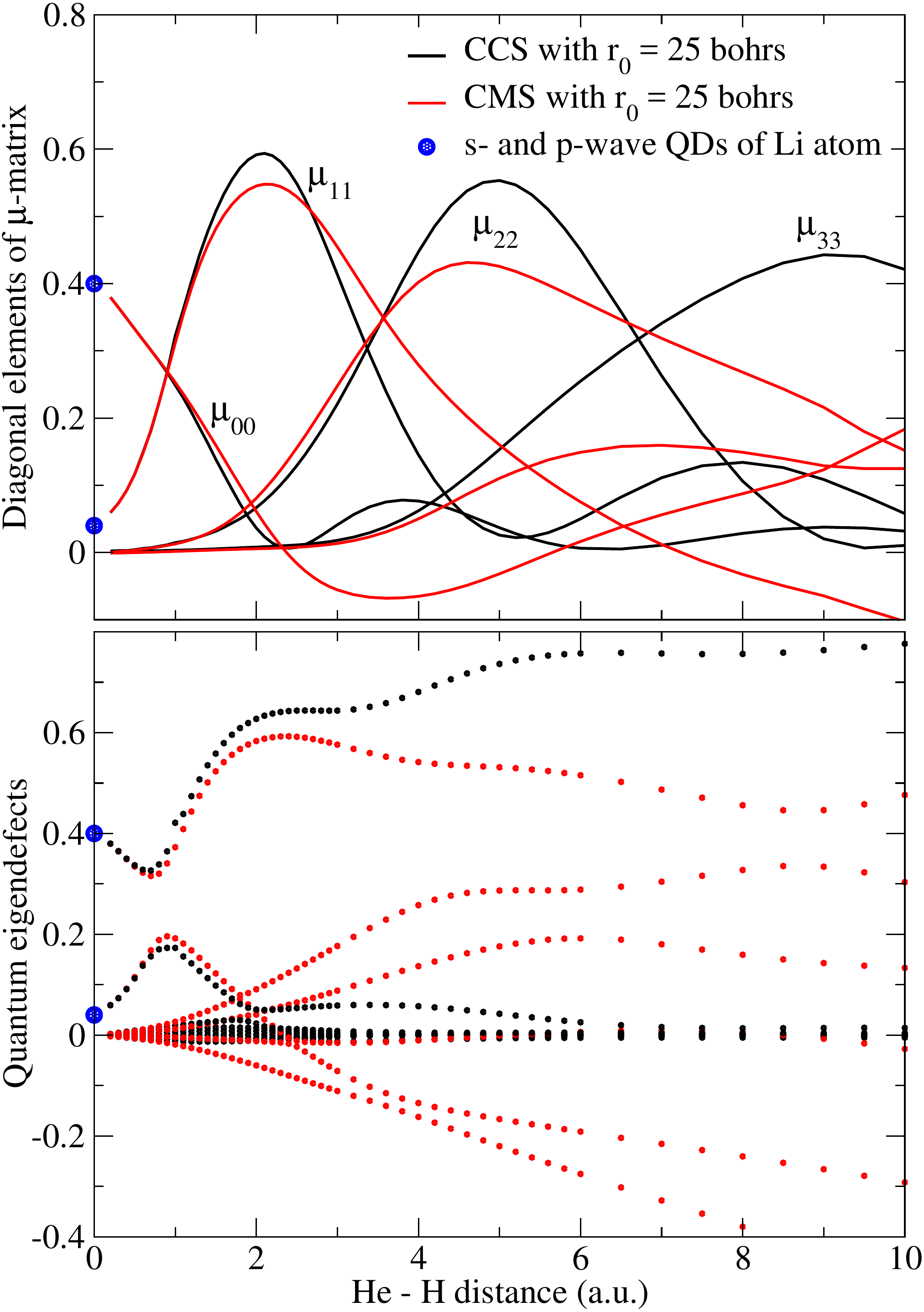}
\end{center}
\caption{\label{fig-QDsig}
The elements of the quantum defect matrix $\mathbf{\mu}$ are shown for $^2\Sigma$-symmetry 
of HeH. The upper panel plots the diagonal elements of the
$\mu$-matrix while the lower panel displays its eigenvalues. CMS stands
for the center-of-mass system while CCS denotes the center-of-charge system.
}
\end{figure}

The quantum defect matrices needed for the present study have been computed using the diatomic UK R-matrix package
\cite{Morgan_Chen_Rm_1997} with the $R$-matrix boundary at $r_0$ = 25 a.u. 
In this calculation, bound electron orbitals are described by the Slater type basis (STO) of
triple-zeta quality (abbreviated as VB2 in Ref.\cite{Ema_Paldus_STO_2003}).
Our zero-energy quantum defects were determined by matching the $R$-matrix,
evaluated at 10 meV, to the Coulomb functions at distances beyond the boundary radius. The possible energy dependence of the 
quantum defects is neglected throughout this study. As is 
mentioned above, Ref.\cite{Takagi_Tashiro_2015} observed that the choice of the
coordinate system origin substantially affects the final DR probabilities. 
In order to address this issue we decided to calculate the two sets of quantum
defects. The first set is evaluated in the center-of-charge system (CCS), while
the second set of quantum defects is computed in the center-of-mass system (CMS).
The comparison of these two approaches is shown in Fig.~\ref{fig-QDsig} for the
$^2\Sigma$-symmetry of the Rydberg electron. The blue circles represent $s$- and
$p$-wave quantum defects of the lithium atom which represents the united atom limit
at $R \rightarrow 0$.

\begin{figure}[tbh]
\begin{center}
\includegraphics[width=0.9\linewidth]{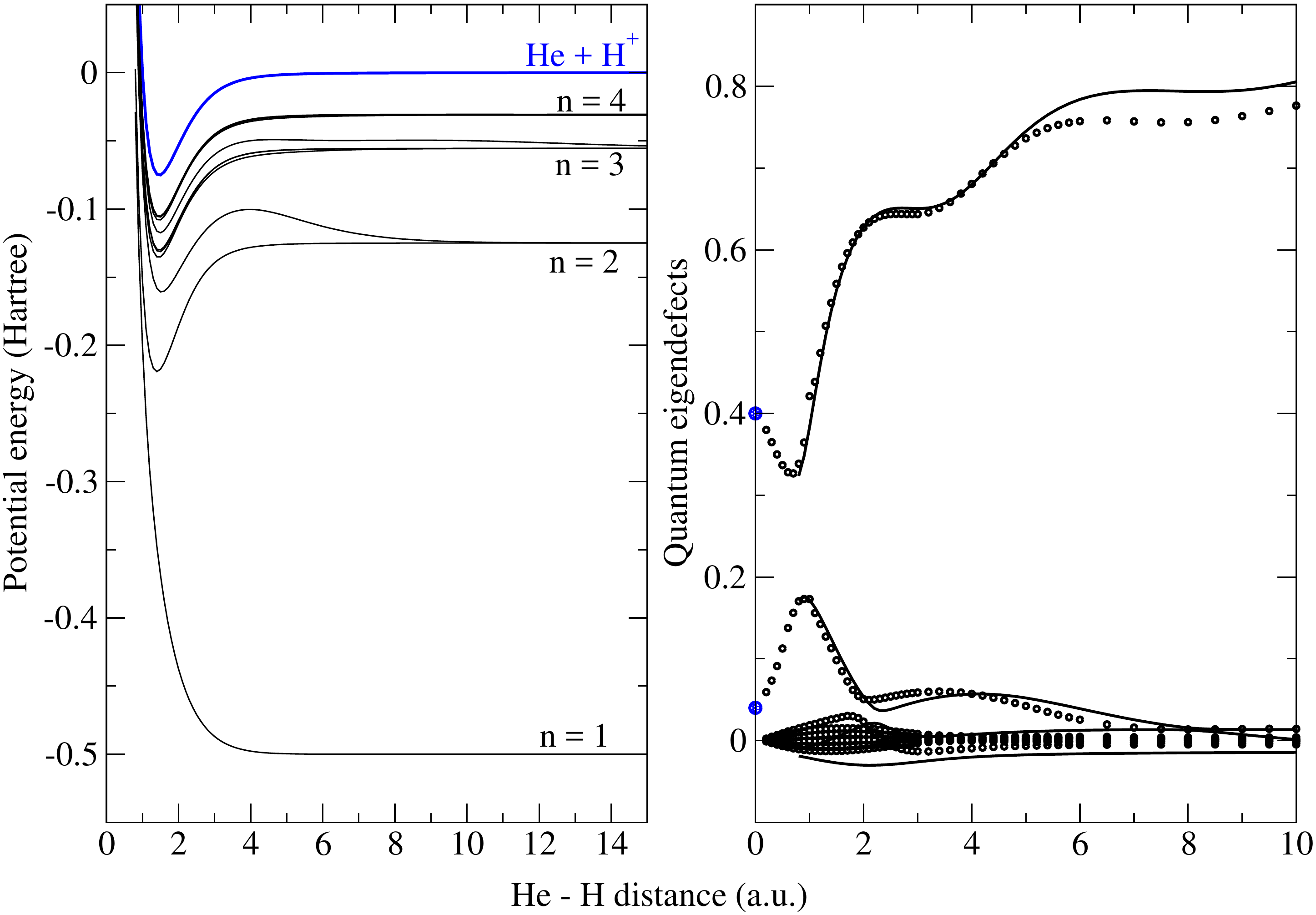}
\end{center}
\caption{\label{fig-pes}
Left panel: Potential energy curves of HeH in $^2\Sigma$-symmetry together with
the ground cation curve. Right panel: quantum eigendefects extracted from $n=4$
Rydberg states of the left panel (dots), lines represent the CCS eigendefects
already shown in Fig.~\ref{fig-QDsig}.
}
\end{figure}

The diagonal elements of the $\mu$-matrix shown
in the upper panel resemble undulations observed previously for LiHe
\cite{Jeung_LiHe_PRA_1999} and for LiH \cite{Altunata_LiH_PRA_2003}. 
These oscillatory features in the potential curves arise from the same mechanism that produces ultra-long-range Rydberg molecules, studied extensively in recent years, and can be understood semi-quantitatively using the Fermi zero-range pseudopotential.  See, for instance, 
\cite{Dolan_Masnou_1981,de_Prunele_1987,Du_Greene_1987,Greene_Dickinson_Sadeghpour_2000,Dickinson_Gadea_2002,Khuskivadze_et_al_2002,Bendkowsky_et_al_2009}. 
The quantum
eigendefects displayed in lower panel of Fig.~\ref{fig-QDsig} confirm the observations
of Takagi and Tashiro \cite{Takagi_Tashiro_2015} that these two sets of
quantum defects differ significantly at larger internuclear distances. On the
other hand there is only one set of quantum eigendefects $\mu_\gamma(R)$ that can be obtained
by inverting the Rydberg diatomic equation (as developed by Mulliken)
\begin{equation}
\label{eq-rydbeg}
U_{n\gamma}(R) = U^+(R) - \frac{1}{2 [n - \mu_\gamma(R)]^2}\;,
\end{equation}
where $U_{n\gamma}(R)$ is the adiabatic potential energy of the Rydberg states and
the potential curve of the cation HeH$^+$ is denoted as $U^+(R)$. The eigendefects
$\mu_\gamma(R)$ are independent of the coordinate system since they can be determined
by a difference of the two potential energy curves. 

Positive energy and negative energy
eigendefects approach each other at zero energy by Seaton's theorem
\cite{Seaton_RPP_1983} and thus our
scattering calculations can be cross-checked by comparing with the Rydberg potential energy curves determined from 
independent quantum chemistry calculations. Such a comparison is shown in 
Fig.~\ref{fig-pes}, where the left panel shows the ground state cation curve 
$U^+(R)$ and the series of $^2\Sigma$-symmetry Rydberg states up to $n$=4. The right
panel (full lines) shows the eigendefects $\mu_\gamma(R)$ extracted through 
Eq.~(\ref{eq-rydbeg}). These results demonstrate that the CCS quantum defects give
correct Rydberg energies while the CMS quantum defects (see Fig.~\ref{fig-QDsig})
differ significantly.

\begin{figure}[th]
\begin{center}
\includegraphics[width=0.6\linewidth]{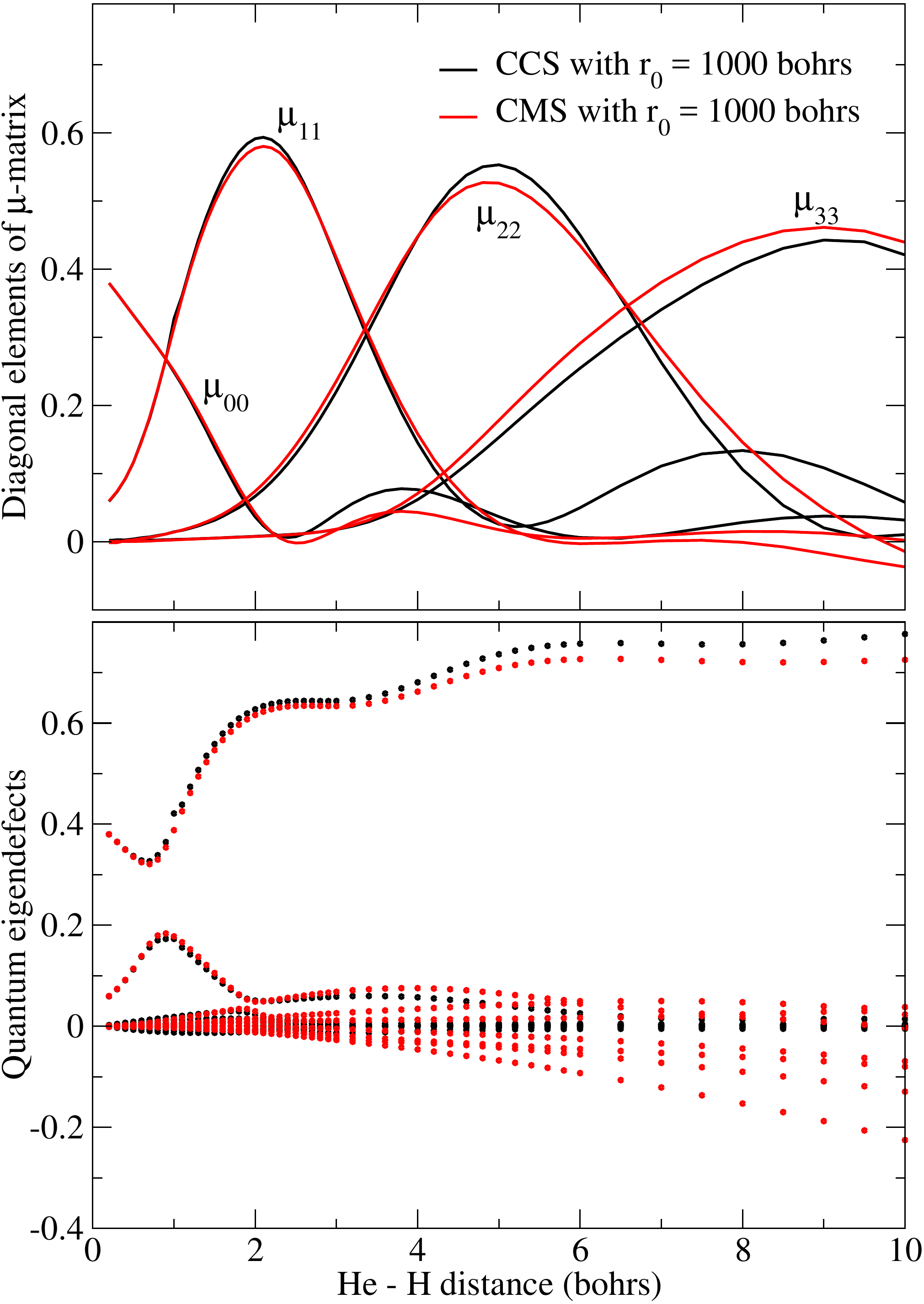}
\end{center}
\caption{\label{fig-qd1k}
Same as Fig.~\ref{fig-QDsig}. Quantum defects were determined by matching at 
$r_0$=1000 a.u.
}
\end{figure}
\begin{figure}[th]
\begin{center}
\begin{tabular}{cc}
\includegraphics[width=0.5\linewidth]{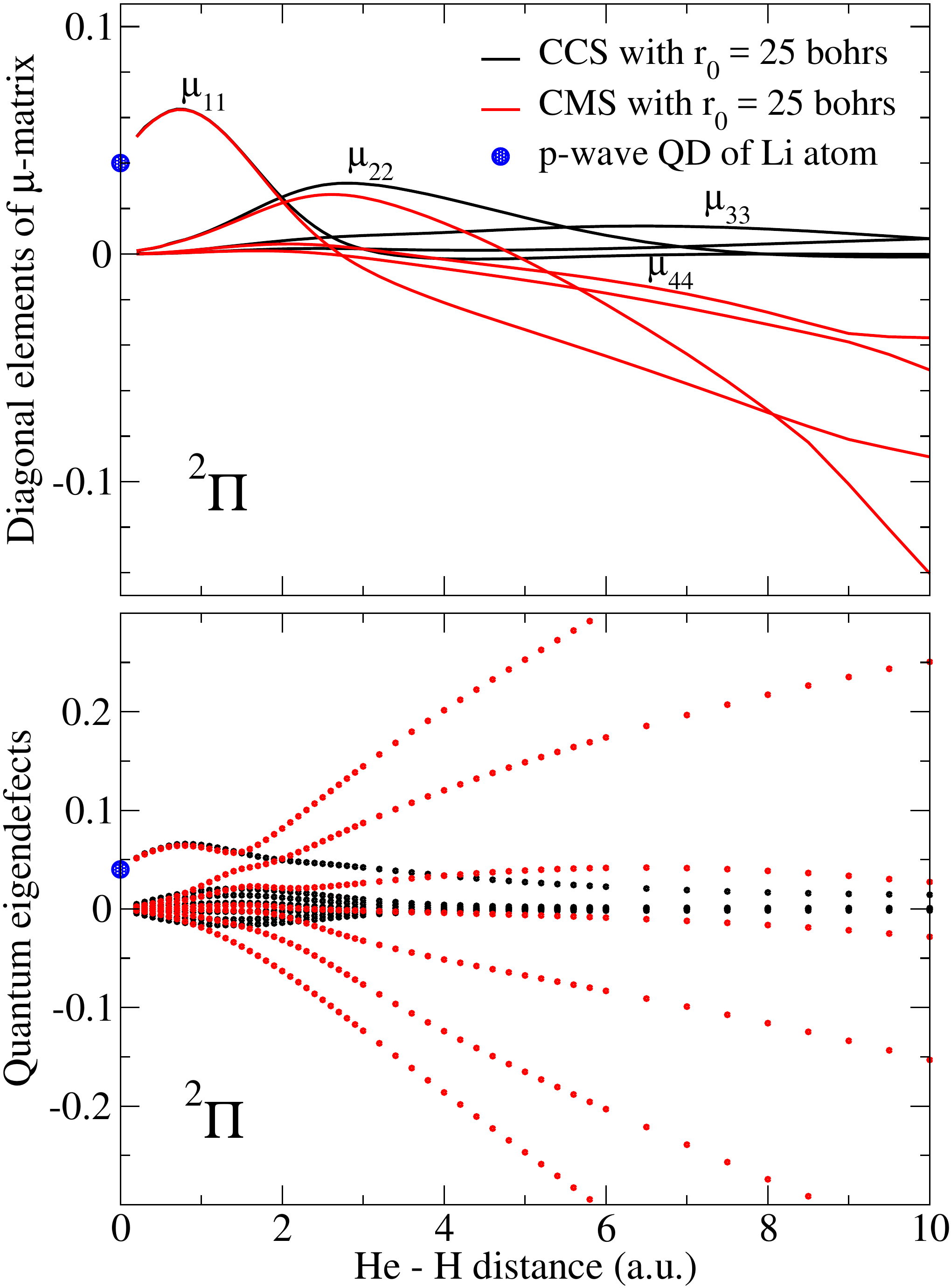} &
\includegraphics[width=0.5\linewidth]{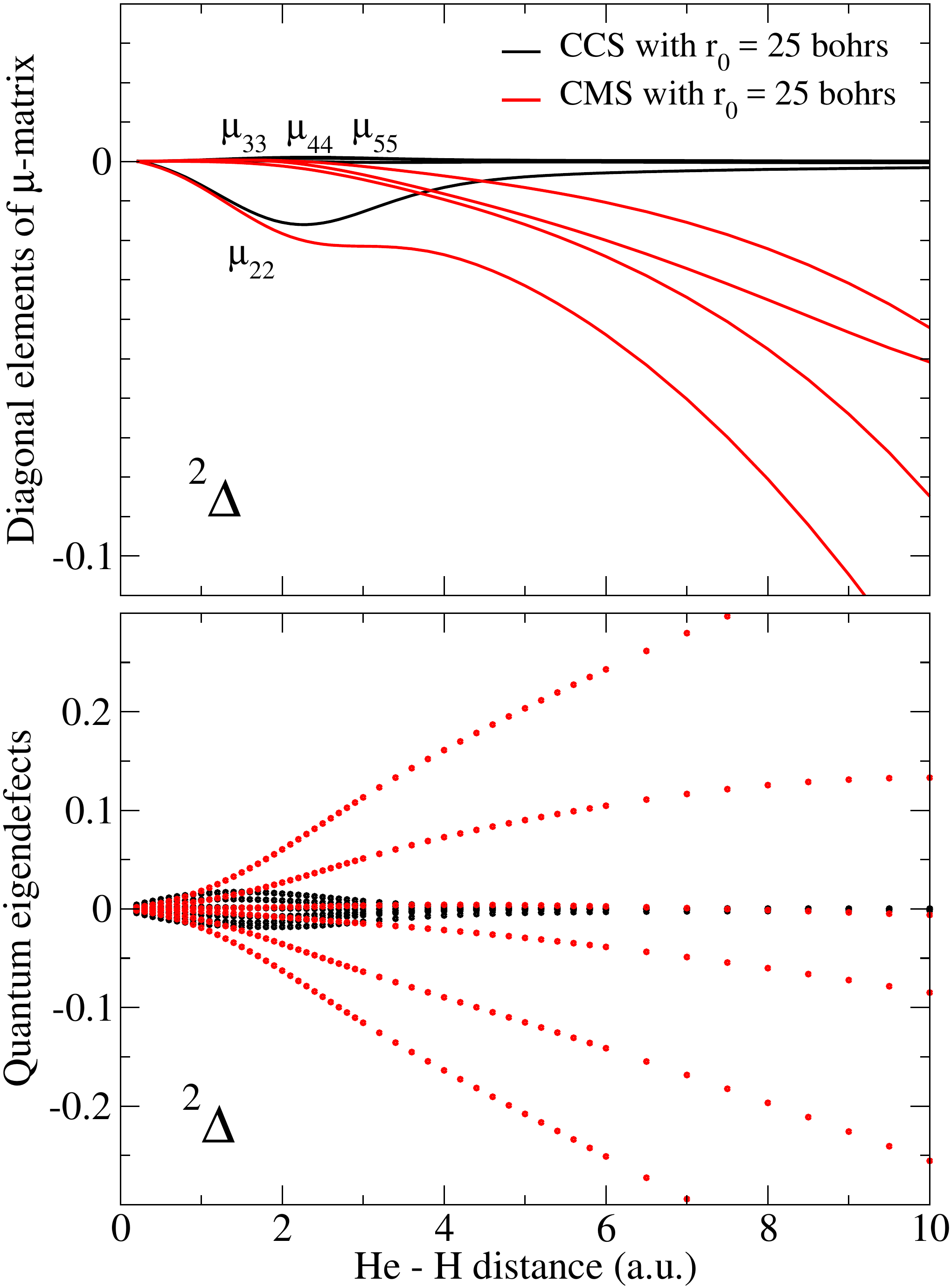} \\
\end{tabular}
\end{center}
\caption{\label{fig-qdpd}
Elements of the quantum defect matrix $\mathbf{\mu}$ for $^2\Pi$ (left panel)
and $^2\Delta$ (right panel) symmetries of HeH.
Upper panels show the diagonal elements of the
$\mu$-matrix while the lower panels display their eigenvalues.
}
\end{figure}
In order to clarify the difference between the CCS and CMS quantum defects, 
the electronic wave function is propagated
to the radial distance $r_0$=1000 a.u. and then the quantum defects matrix
is determined. In the case of the CCS the asymptotic potential has a form of the Coulomb
field and the quantum defects determined at $r_0$=1000 a.u. are identical with
those determined at $r_0$=25 a.u. (shown in Fig.~\ref{fig-QDsig}).
However, in the case of the CMS propagation the asymptotic potential also
contains a dipolar interaction that gives an additional phase (quantum defect) gain
together with an enhanced coupling of the partial waves involved.
Resulting comparison of the two models (CCS and CMS) propagated to $r_0$=1000 a.u. is
shown in Fig.~\ref{fig-qd1k}. The two sets of quantum defects become very similar.
Small differences at larger $R$ can be attributed to 
the dipole moment coupling of partial waves in the outer region.
The angular space presently limited by $l_{max} = 5$ may be insufficient to correctly
describe the net phase gain of the lowest partial waves.

To summarize, the only differences among the quantum defects determined at a finite
radius $r_0$ and for different centers of the origin can be attributed to a phase gain
in the outer dipole field. This phase gain is missing for those choices of the origin
that produce the asymptotic dipole potential. These conclusions constitute a theoretical
problem that appears during the frame transformation technique and it will be discussed
later.

For the sake of completeness we also present the quantum defects for $^2\Pi$ and $^2\Delta$ 
symmetries, as shown in Fig.~\ref{fig-qdpd}. The missing dipole propagation in the CMS
case gives a strong effect because the $^2\Pi$ and $^2\Delta$ quantum defects are smaller
when compared to the $^2\Sigma$ symmetry.

\section{\label{sec-FTs}Frame transformations}

The role of the frame transformation (FT) here is to connect the total wave function
(electronic and nuclear) in two different electronic spatial volumes. The first (inner)
spatial region is defined by the interior of a sphere (of the radius $r_0$) enclosing the HeH$^+$
system. In the second (outer) region where $r \geq r_0$, we assume that 
the interaction between the electron and the cation is purely Coulombic.
In the two spatial regions we use different, but physically appropriate, quantization
schemes for the nuclear degrees of freedom. 
Inside the sphere ($r \leq r_0$) the Born-Oppenheimer approximation 
often holds, and the colliding electron moves in the potential of fixed nuclei, 
whereby the nuclear positions are considered as ``good quantum numbers" describing this part of the electron motion.
On the other hand the asymptotic channels outside the sphere are defined
by the rovibrational states of the cation that are coupled by the
electronic part of the wave function expressed by Coulomb functions.
Therefore, the radius $r_0$ should be small enough so the Born-Oppenheimer
approximation is accurate inside there sphere. This requirement,
when combined with our conclusions on the CMS and CCS quantum defects,
represents a dilemma for determining the most appropriate FT distance $r_0$:
\begin{itemize}
\item [a.] At $r_0$ = 25 Bohr radii the CMS quantum defects behave wildly
(Figs. \ref{fig-QDsig} and \ref{fig-qdpd}), they are likely not converged
with respect to the number of partial waves and their corresponding eigendefects do not
reproduce energies of the bound Rydberg states. On the other hand, the CCS
quantum defects diminish quickly with the higher angular momentum $l$ and its projection
$\Lambda$. They also satisfactorily reproduce the bound state energies via Seaton's
theorem (Fig.~\ref{fig-pes}). However, a direct application of the CCS defects is
complicated by physics - the molecule does not rotate around the center of charge. 
In reality the electron collides with a rotating dipolar cation.
\item[b.] At $r_0$ = 1000 Bohrs the differences between the two sets of quantum defects
disappear and the bound state energies are also well reproduced by the CMS
quantum defects that are more physical to consider for the rotational FT.
An inherent problem with this approach stems from the limited accuracy of the Born-Oppenheimer approximation
inside a volume of such large radius.
\end{itemize}

It is clear that both of the available options bring a certain level of an approximation into
the computational model. Corresponding inaccuracies can hardly be estimated beforehand and therefore,
we have chosen to carry out the FT for both,  CMS and CCS quantum defects, in order to critically compare resulting inelastic and DR rates. 

\subsection{\label{ssec-RVstates}Rovibrational states}

The rovibrational nuclear states of the HeH$^+$ cation define the asymptotic electron escape
channels and they have the form of \cite{Chang_Fano_1972} (Hund's case b)
\begin{equation}
\chi_{\nu j }(R,\hat{\bm{R}}) = \phi_{\nu j}(R) Y_{j M-m}(\hat{\bm{R}})\;,
\end{equation}
where $M=m+m_j$ is the projection of the total angular momentum $\bm{J}=\bm{l}+\bm{j}$ in the LAB
frame, $\bm{j}$ is the angular momentum describing the molecular rotations, $\bm{l}$ is the
orbital angular momentum of the colliding electron, and $\nu$ indexes the molecular 
vibrational functions. 
\begin{figure}[ht]
\begin{center}
\includegraphics[width=0.7\linewidth]{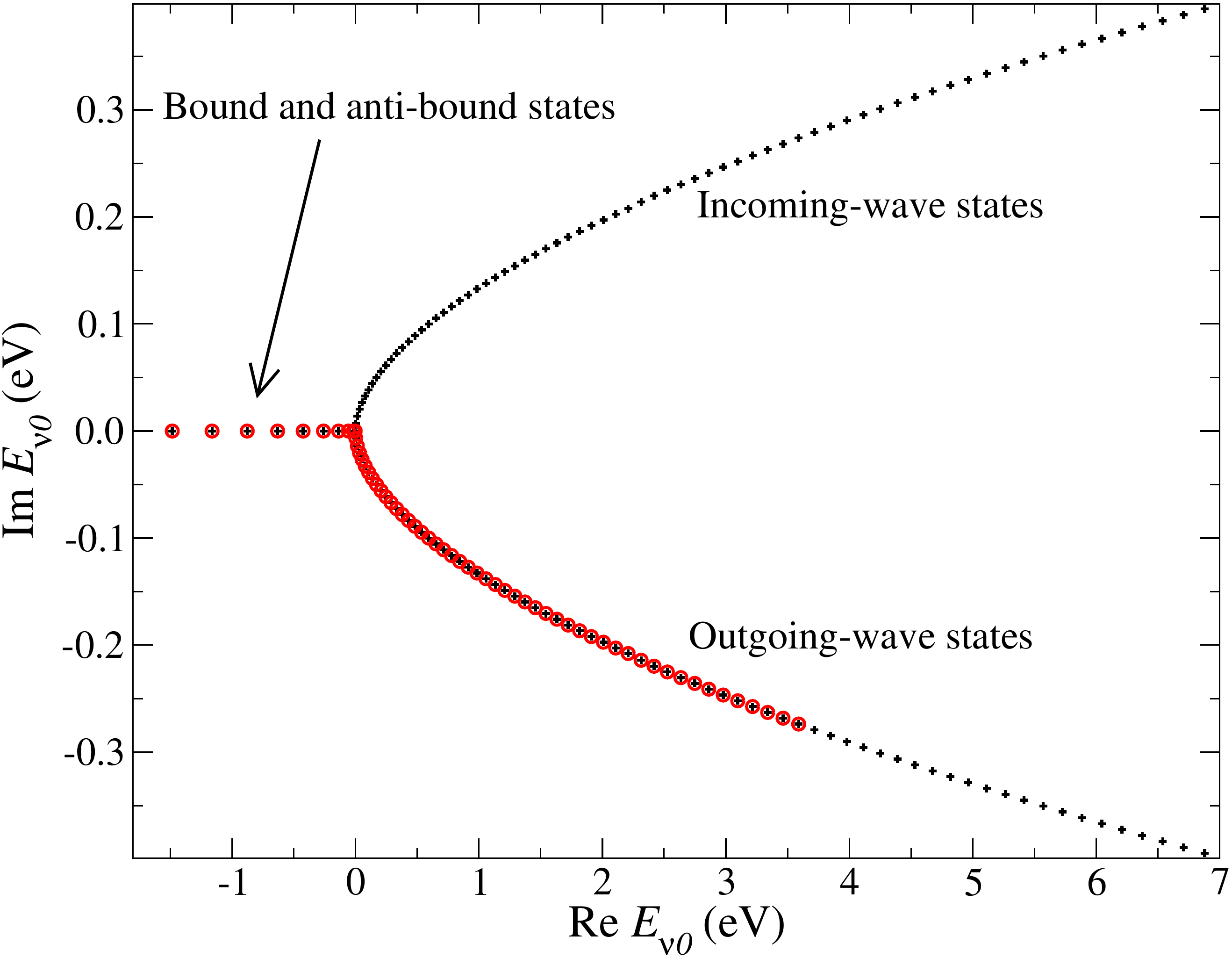}
\end{center}
\caption{\label{fig-cplane}
Distribution of the energy eigenvalues $E_{\nu j} = K^2_{\nu j}/2\mu$ (Eq.~\ref{eq-Schro-nuc})
in the complex plane for $j=0$. Circles denote the 60 states used in the present calculations.
}
\end{figure}
The rovibrational states 
$\phi_{\nu j}(R)$ and corresponding energies were obtained by solving 
nuclear Schr\"{o}dinger equation
\begin{equation}
\label{eq-Schro-nuc}
\left[ - \frac{\mathrm{d}^2}{\mathrm{d}R^2} 
+ 2 M_r U^+(R) + \frac{j(j+1)}{R^2}
- K^2_{\nu j} \right] \phi_{\nu j}(R) = 0 \;,
\end{equation}
with Siegert
\cite{Siegert_1939,Tolstikhin_2007}
boundary conditions at the origin and at $R_0$~=~10~a.u.
\begin{equation}
\label{eq-vib-SPS2}
\phi_{\nu j}(0) = 0; \;\;\;\;\; \phi_{\nu j}(r) \sim R h^{(1)}_j(K_{\nu j}R)
\hbox{  for  } R \geq R_0\;,
\end{equation}
where $h^{(1)}_j$ is the spherical Hankel function of the first kind. In other
words, the boundary condition at $R_0$ selects only those solutions, whose
asymptotic momentum $K_{\nu j}$ coincides with the energy eigenvalues
$E_{\nu j}=K_{\nu j}^2/2\mu$, leading to a form of continuum discretization as shown
in Fig.~\ref{fig-cplane}. The circled states denote a subset of 60 Siegert states that
resulted in converged results for electron collision energies up to 2.5~eV.

The cation curve $U^+(R)$ employed in Eq.~(\ref{eq-Schro-nuc}) (also shown by the blue line
in Fig.~\ref{fig-pes}) was computed by the full CI method using Dunning's aug-cc-pV5Z basis
\cite{Woon_Dunning_1994}
as implemented in Molpro 2012. The reduced mass $M_r = 1467.28$ (for $^4$HeH$+$)
was taken from
Ref.~\cite{Coxon_Hajigeorgiou_1999} (denoted as $\mu_{eff}$ in Tab.~3). Resulting
rovibrational energies are in a very good agreement (within 2 cm$^{-1}$) with accurate
calculations of Pachucki and Komasa 2012 \cite{Pachucki_Komasa_2012} for $\nu \leq 5$ and
$j \leq 12$ listed in Tab. III of Ref.~\cite{Pachucki_Komasa_2012}.

\subsection{\label{ssec-RVFT}Rovibrational frame transformation}

We implement the rovibrational FT as a two-step procedure. The first step is the
vibrational frame transformation
\cite{Chang_Fano_1972,Greene_Jungen_PRL_1985}, later modified
\cite{Hamilton_Greene_PRL_2002} to account for the modified Siegert state
orthonormality relation, via
\begin{equation}
\label{eq-vft}
S^\Lambda_{l\nu,l'\nu'}(j,j') = \int_0^{R_0}\!\!\mathrm{d}R \phi_{\nu j}(R) 
S^\Lambda_{l l'}(R) \phi_{\nu'j'}(R) + i \frac{\phi_{\nu j}(R_0)
S^\Lambda_{l l'}(R_0)\phi_{\nu'j'}(R_0)}{K_{\nu j}+K_{\nu'j'}}\;.
\end{equation}
The second step is accomplished by the rotational frame transformation
\cite{Chang_Fano_1972}
\begin{equation}
\label{eq-rft}
S^{J\eta}_{l\nu j,l'\nu'j'} = \sum_\Lambda U^{J \eta l}_{j \Lambda}
S^\Lambda_{l\nu,l'\nu'}(j,j') U^{J \eta l'}_{j' \Lambda}\;,
\end{equation}
where 
$\eta = (-1)^{l+j}$ is the total parity of the neutral
system. It is worth to note that in the present dissociative recombination
results the contribution of the odd parity cross section is negligible
(less than 0.5\%) because only $\Lambda > 0$ contributes to
the $\eta = -1$ symmetry
\cite{Chang_Fano_1972} and the small $^2\Pi$ and $^2\Delta$ quantum defects,
shown in Fig.~\ref{fig-qdpd}, seem
to have a weak impact on the dissociation dynamics.

Due to the Coulomb nature of the asymptotic potential, the matrix
$S^{J\eta}_{l\nu j,l'\nu' j'}$ exhibit only a very weak energy dependence, which
is completely neglected in the present study. Moreover, it cannot be 
regarded as the inelastic $S$-matrix of the scattering theory as it represents
merely a set of coefficients coupling rovibrational channels with the
electronic wave functions for which the physical asymptotic boundary conditions have not yet been
applied, namely in the closed channels. The application of the
electronic boundary conditions in the MQDT closed-channel elimination step leads to the physically relevant
$S$-matrix, which of course is defined only in the open-channel space \cite{Orange_review}
\begin{equation}
\label{eq-elimc}
\bm{S}^{\mathrm{phys}} = \bm{S}^{oo} - \bm{S}^{oc}\left[
\bm{S}^{cc}-e^{-2 i \bm{\beta}(E)} \right]^{-1} \bm{S}^{co} \;,
\end{equation}
where the superscripts $o$ and $c$ denote open and closed sub-blocks in the
unphysical $S$-matrix $S^{J\eta}_{lj\nu,l'j'\nu'}$, respectively.
The diagonal matrix $\bm{\beta}(E)$ describes effective Rydberg quantum
numbers with respect to the close-channel thresholds:
\begin{equation}
\label{eq-elimb}
\beta_{ij} = \frac{\pi}{\sqrt{2(E_i-E)}}\delta_{ij}\;.
\end{equation}
%

\section{\label{sec-rovib}Rovibrational excitation}

The rovibrationally inelastic cross section is computed from the physical
$S$-matrix $S^{\mathrm{phys}, J\eta}_{l\nu j,l'\nu'j'}$ as
\cite{Chang_Fano_1972,Morrison_greenbook1}
\begin{equation}
\label{eq-ICS-ine}
\sigma_{(\nu'j')\rightarrow (\nu j)}(\varepsilon_{\nu' j'}) =
\frac{\pi}{2\varepsilon_{\nu' j'}(2j'+1)}
\sum_{J \eta l l'} \left(2J+1\right) \left|T^{J\eta}_{l\nu j,l'\nu'j'}\right|^2\;,
\end{equation}
where $\varepsilon_{\nu'j'}=E-E_{\nu'j'}$ is the incident electron collision energy 
and the state-to-state scattering amplitudes $T^{J\eta}_{l\nu j,l'\nu'j'}$ are defined by 
\begin{equation}
\label{eq-Tmat}
T^{J\eta}_{l\nu j,l'\nu'j'} = S^{\mathrm{phys}, J\eta}_{l\nu j,l'\nu'j'} -
\delta_{ll'}\delta_{\nu\nu'}\delta_{jj'}\;.
\end{equation}
The corresponding inelastic rate coefficient $\alpha(\varepsilon)$ is obtained by multiplying
the cross section by velocity, and thermally averaging it when necessary:
\begin{equation}
\label{eq-rate1}
\alpha(\varepsilon) =
\sqrt{2 \varepsilon}\; \sigma(\varepsilon)\;.
\end{equation}

\begin{figure}[tbh]
\begin{center}
\includegraphics[width=1.0\linewidth]{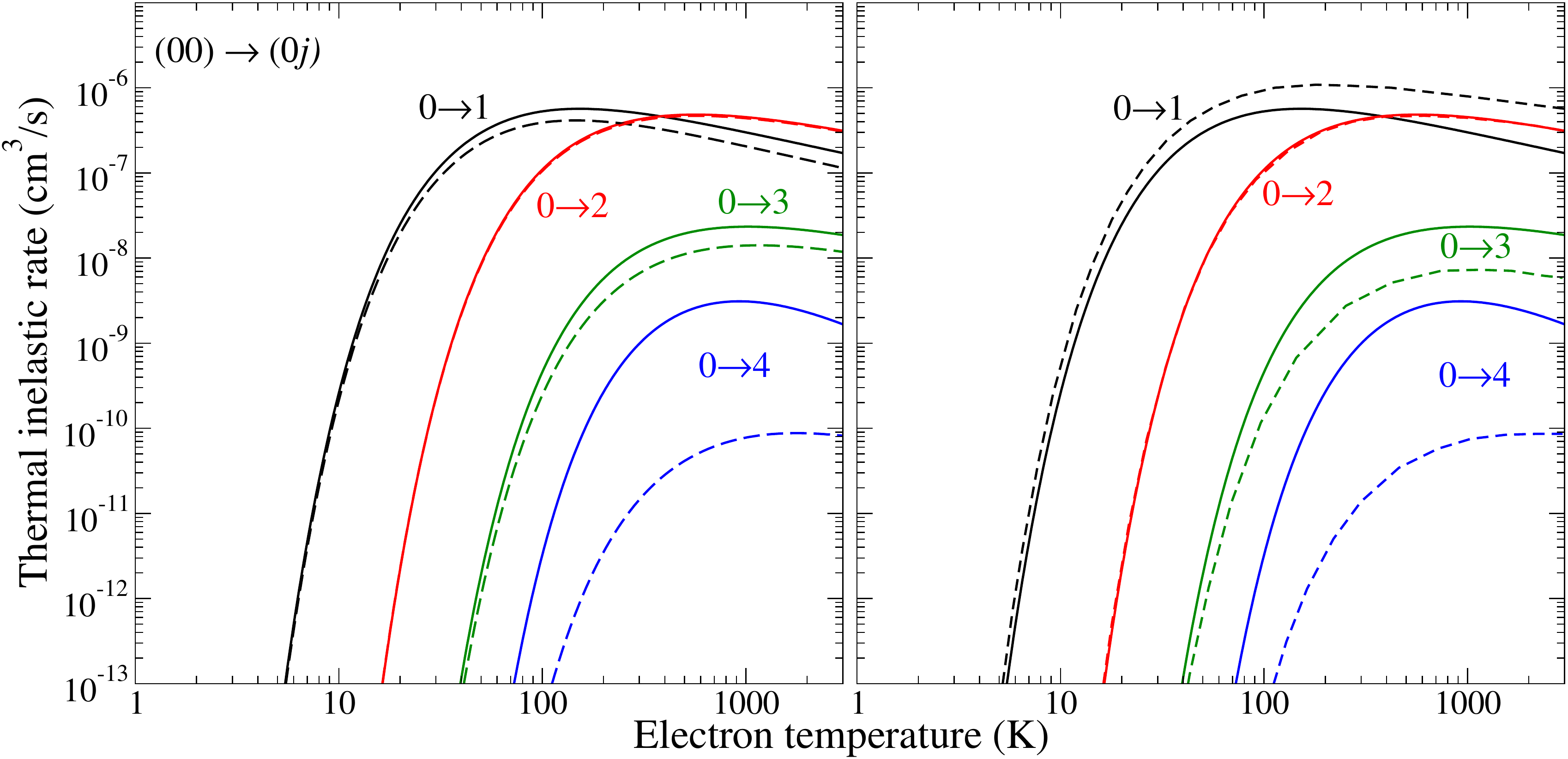}
\end{center}
\caption{\label{fig-rvt}
Thermal rates of rotational excitation of HeH$^+$
from the rovibrational ground state $(\nu j) = (00)$. The left panel
shows comparison of the present center-of-mass (full curves) and
and center-of-charge (broken curves) results. The right panel
compares the present center-of-mass results with recent CMS calculations
of Hamilton {\it et al} 2016 \cite{Hamilton_Tennyson_HeH_2016}. Note
that in both panels the broken and full curves overlap for the
0$\rightarrow$2 excitations.
}
\end{figure}

The discussion of our results begins with a comparison of the rotational excitation rates of the HeH$^+$ cation
computed in the center-of-mass and center-of-charge systems. Both rates
are shown in the left panel of Fig.~\ref{fig-rvt}. It is not
surprising that the non-zero dipole moment in the center-of-mass system
visibly influences the 0$\rightarrow$1 and 0$\rightarrow$3 transitions.
The transition 0$\rightarrow$4 is also strongly impacted by the choice
of the frame of reference (or a distance $r_0$ at which the rotational
FT is carried out), however, the cross sections are already quite low.

The present center-of-mass results are also compared to recent calculations
of Hamilton {\it et al} 2016 \cite{Hamilton_Tennyson_HeH_2016}, as displayed
in the right panel of Fig.~\ref{fig-rvt}. Aside from the 0$\rightarrow$1 transition,
the remaining results of Hamilton {\it et al} 2016 strongly resemble the present
CCS results (and the same as the CMS results with the electronic calculation 
performed out to a large radius, $r_0$ = 1000 a.u.).
This similarity can be explained by the differences in the
two theoretical models used to compute the inelastic rates:
\begin{itemize}
\item[(i)] In the present calculations the scattered electron is described at
the sphere with $r_0$ = 25 a.u. with partial waves $l \leq 5$. While the rotational
FT at this distance is quite accurate, the present CMS results neglect the external dipole field
for $r>r_0$.
\item[(ii)] In the calculations of Hamilton {\it et al} 2016
\cite{Hamilton_Tennyson_HeH_2016} the R-matrix radius is chosen at $r_0$ = 13 Bohr radii and
the partial wave expansion is limited by $l \leq 4$. However, the LAB frame propagation
for $r > r_0$ is accounted for by the Coulomb-Born closure technique
\cite{Rabadan_Tennyson_1998}
that is designed to add, incoherently, the missing cross-section contribution associated with higher
partial waves in the LAB frame domain. Such procedure is dominant for the
$\Delta j = \pm 1$ transitions and it can also be viewed as effectively moving the frame transformation
radius $r_0$ to infinity. This can explain the good agreement with the present center of mass
results with FT at $r_0$ = 1000 Bohrs, aside from the 0$\rightarrow$1 transition. 
For this transition it is difficult to decide which of the two theoretical models is more accurate.
\end{itemize}

\begin{figure}[th]
\begin{center}
\includegraphics[width=1.0\linewidth]{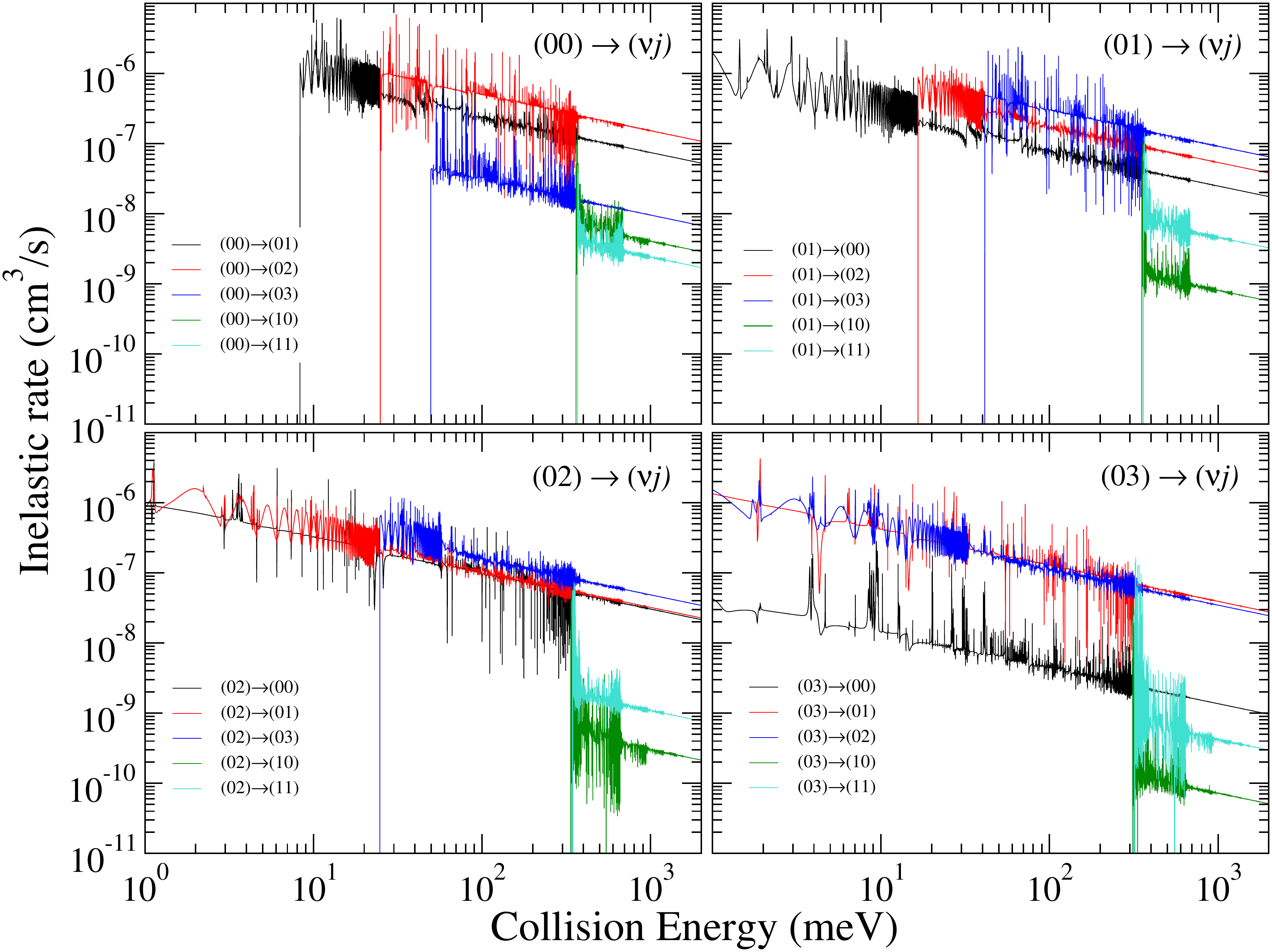}
\end{center}
\caption{\label{fig-rv0}
Computed rates for the inelastic rovibrational transitions of HeH$^+$, starting from the initial vibrational ground
state $\nu'$ = 0.}
\end{figure}

The data displayed in Fig.~\ref{fig-rvt} are a result of a thermal Boltzmann convolution
(see for example \cite{Curik_Greene_MP_2007}) of the energy-dependent inelastic
rates. Such rates are shown in Fig.~\ref{fig-rv0} for the four lowest rovibrational
initial states. Observe that all the data presented in Fig.~\ref{fig-rv0}
share the importance of the two dominant
transitions $\Delta j = \pm 1$ and $\Delta j = \pm 2$.

\section{\label{sec-DR}Dissociative recombination}

The physical $S$-matrix (\ref{eq-elimc}) is the result of the ``closed-channel elimination"
technique of MQDT that applies proper electronic boundary condition in the closed channels, i.e.
exponentially decaying behavior for the components of the electronic wave function
in the closed channels.  The present formulation does not present this wave function in a
form that would also enforce proper nuclear boundary conditions in order to obtain the full scattering matrix that includes explicit atom-atom or ion-pair dissociation channels that would yield
the DR cross section directly, as was done in the previous theoretical studies
\cite{Guberman_HeH_1994,Takagi_Tashiro_2015}. Instead, we use the fact that the
electronic $S$-matrix (\ref{eq-elimc}) in the nuclear basis of the complex 
outgoing-wave Siegert states is subunitary. Following the original idea of Hamilton and Greene
2002 \cite{Hamilton_Greene_PRL_2002}, we identify the resulting loss of electronic flux
with the only physical mechanism (in the Hamiltonian utilized which omits the radiation field coupling) that can lead to the loss of electrons, namely the dissociative
recombination events. This physical idea leads to a mathematical expression for the DR cross
section as a function of the incident energy $\varepsilon_{\nu'j'}$ for the target ion
in an initial state defined by $(\nu'j')$:
\begin{equation}
\label{eq-ICS-DR}
\sigma^{DR}_{\nu'j'}(\varepsilon_{\nu'j'}) =
\frac{\pi}{2\varepsilon_{\nu'j'}(2j'+1)}
\sum_{J \eta l'} \left(2J+1\right) \left(
1 - \sum_{l \nu j}
S^{\mathrm{phys}, J\eta}_{l\nu j,l'\nu'j'}\; 
S^{+\mathrm{phys}, J\eta}_{l'\nu'j',l\nu j} \right)\;.
\end{equation}
The recombination rate is then connected with the DR cross section using
the simple expression (\ref{eq-rate1}).

\subsection{\label{ssec-dr-cmsccs}Center-of-charge vs. center-of-mass}

\begin{figure}[th]
\begin{center}
\includegraphics[width=0.9\linewidth]{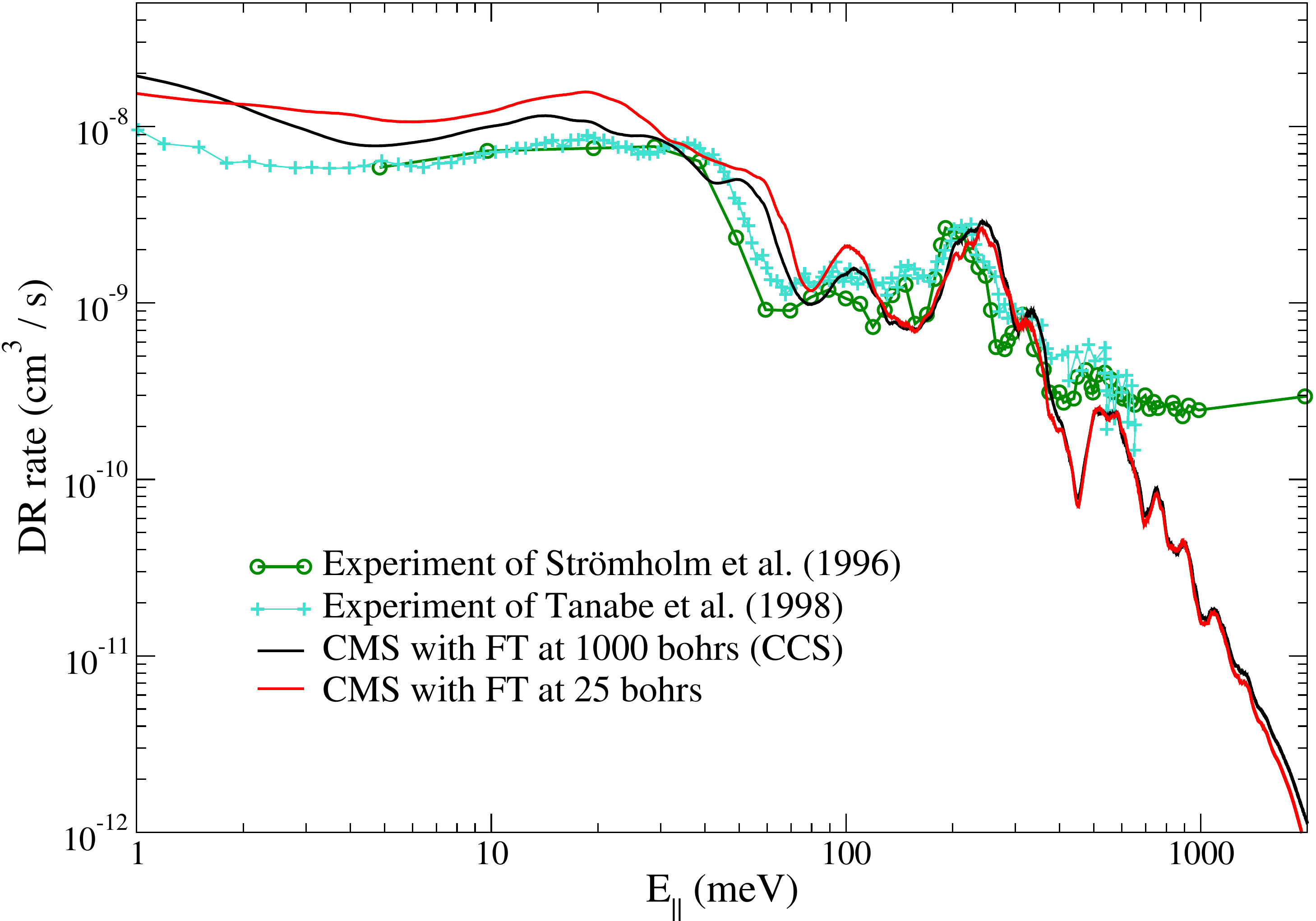}
\end{center}
\caption{\label{fig-cmsccs}
The DR rates are compared for two calculations with quantum defect matrices determined either in the center-of-charge
(CCS) or center-of-mass (CMS) reference systems. The corresponding colors of the quantum defects
shown in Figs.~\ref{fig-QDsig} and \ref{fig-qdpd} are used. 
Results are averaged over the Boltzmann population
of the cation initial states (at 800K) and over the electron beam energy spread appropriate to a typical current generation storage ring experiment:
$\Delta E_{||}$ = 0.1 meV and $\Delta E_{\perp}$ = 10 meV.
The absolute data of
Str\"{o}mholm {\it et al.} (1996) \cite{Stromholm_Larsson_HeH_1996} are displayed
as green circles, while the scaled data of Tanabe
{\it et al.} (1998) \cite{Tanabe_Takagi_HeH_1998} are shown as turquoise crosses.
}
\end{figure}

The CCS and CMS results are compared
in Fig.~\ref{fig-cmsccs}. Our rapidly oscillating results full of dense resonance regions are
convolved over the electron beam energy spreads of
$\Delta E_{||}$ = 0.1 meV and $\Delta E_{\perp}$ = 10 meV, estimated in
the previous experiments
\cite{Stromholm_Larsson_HeH_1996,Tanabe_Takagi_HeH_1998}.
Evidently the difference between the two theoretical approaches is
quite small at lower energies (below 100 meV) and it almost disappears
at the collision energies above 100 meV.
This finding is in clear disagreement (by about 2 orders
of magnitude) with the difference demonstrated by Takagi and Tashiro 
\cite{Takagi_Tashiro_2015} (see Figs.~2 and 3 in the reference).
Moreover, the difference between the center-of-mass and the center-of-charge
results is so small that the experimental data can hardly help to decide which
of the two models is more accurate.
We conclude that the 
center-of-charge results represent the experiments slightly better, and the numerical
aspects of our calculations are also under better control in this case
(especially the convergence with the number of partial waves included). Therefore,
in the following only the center-of-charge DR results are presented.

\subsection{\label{ssec-dr-isotopes}DR for the other isotopologues}

\begin{figure}[ht]
\begin{center}
\includegraphics[width=1.0\linewidth]{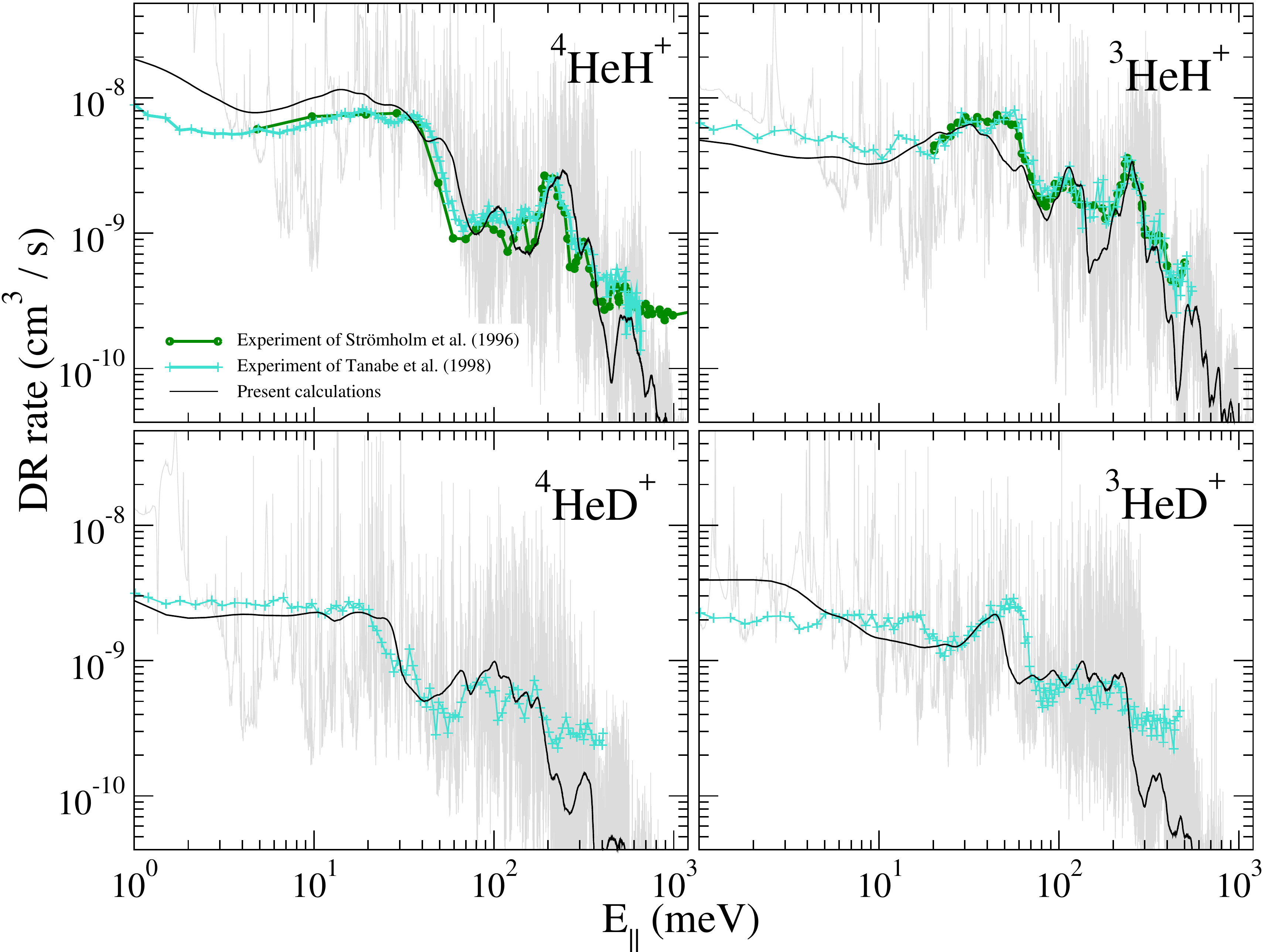}
\end{center}
\caption{\label{fig-isotopes}
Computed and measured DR rate for four different isotopologues of
HeH$^+$ cation. Absolute data of
Str\"{o}mholm et al. (1996) \cite{Stromholm_Larsson_HeH_1996} 
are displayed as green circles, while the scaled data of Tanabe
et al. (1998) \cite{Tanabe_Takagi_HeH_1998} are shown as turquoise crosses.
In the upper two panels the data of Tanabe
et al. (1998) are scaled to the experiment of
Str\"{o}mholm et al. (1996). In the lower two panels the experiment
is scaled arbitrarily to the present calculations shown by the black
line.
The data in grey represent calculations without the averaging over
electron beam energy distributions (Boltzmann population of the initial
rovibrational states of $^4$HeH$^+$ is still accounted for).
}
\end{figure}

The change of the atomic isotopes in the HeH$^+$ cation has no impact
on the electronic part of our CCS calculations that are described in Section~\ref{sec-QDs}. 
This change is reflected
only in the Siegert states spectrum via modification of the reduced
mass $M_r$ in Eq.~(\ref{eq-Schro-nuc}).
Since the reduced mass of
$^4$HeH$^+$ taken from Tab.~3 of Ref. \cite{Coxon_Hajigeorgiou_1999}
in combination with the presently calculated cation curve resulted in
very accurate rovibrational levels, the effective masses listed in 
Table.~3 of Ref. \cite{Coxon_Hajigeorgiou_1999} are also adopted for the
other isotopologues. Rovibrational levels calculated with the
Siegert equations (\ref{eq-Schro-nuc}) and (\ref{eq-vib-SPS2})
yield rovibrational transitions that can be compared with the
experimental data (Tab.~6 of
Ref.~\cite{Coxon_Hajigeorgiou_1999}). The agreement is again
very good with differences within 1 cm$^{-1}$ for the first four
R-lines listed for the each isotopologue.
Resulting DR rates are shown in Fig.~\ref{fig-isotopes} for all the four
isotopologues.

The calculated DR rates generally reproduce the structures seen in the available
experiments. The computed data for the other isotopologues share
the high-energy deficiency visible already for $^4$HeH$^+$ and also in the
previous calculations of Haxton and Greene 2009
\cite{Haxton_Greene_HeH_2009}.
This deficiency is represented by lower computed DR rates 
for higher collision energies
(above several hundreds of meV). Such discrepancy can presumably be explained by
the so-called "toroidal correction" that is not included in the present study
\cite{Slava_Greene_PRA_2005}. The toroidal correction accounts for collision
events at the edges of the collision area in which the electron and ion
beam are merged in the storage rings. The relative collision energies in those edges
depend on the velocity vectors of the merging beams. That dependence
 effectively widens the velocity spreads of the beams and it can also
result in an additional background rate brought from the higher-energy collision
events.

\subsection{\label{ssec-dr-ss}Initial state-dependent DR}

\begin{figure}[ht]
\begin{center}
\includegraphics[width=1.0\linewidth]{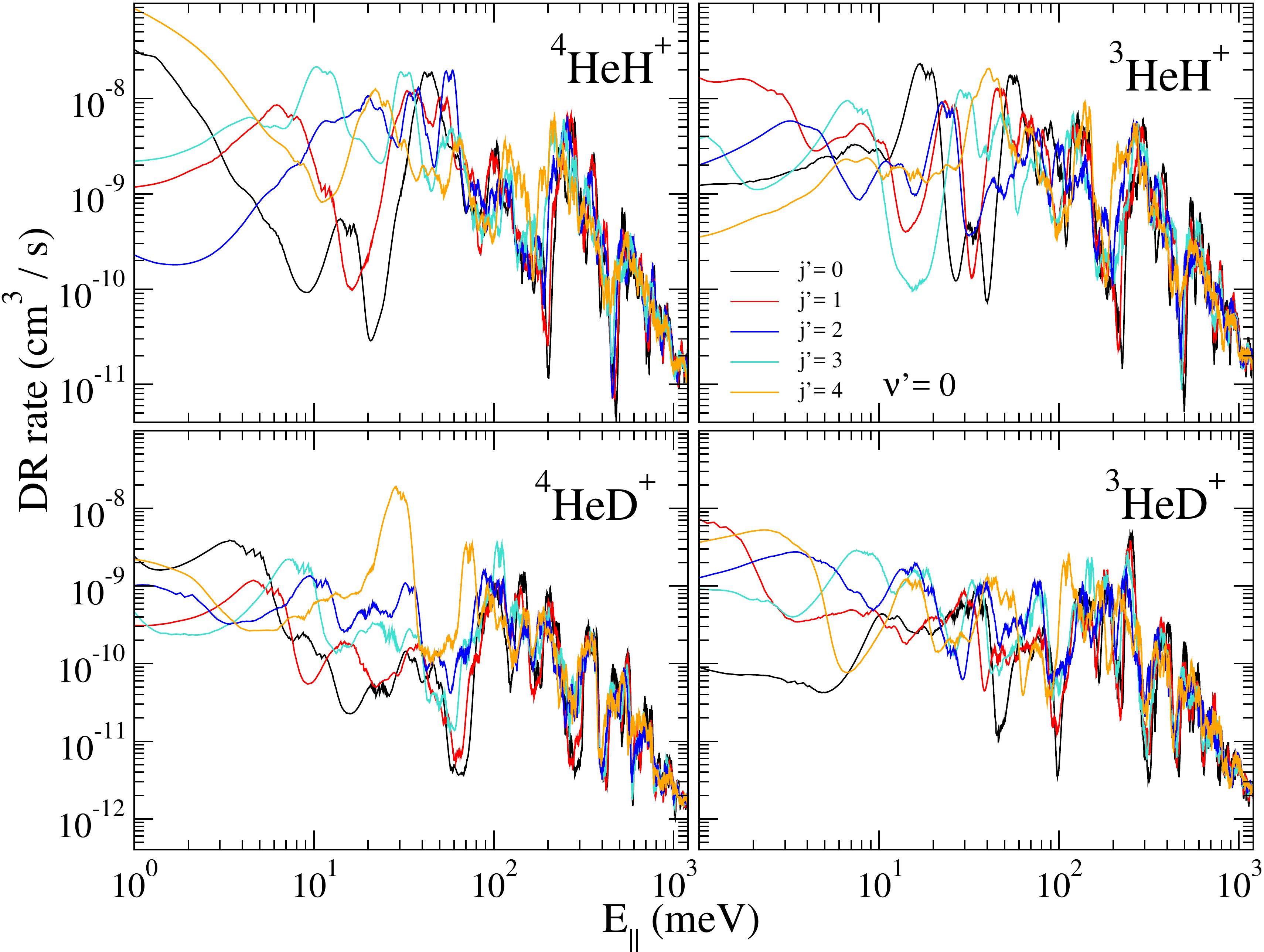}
\end{center}
\caption{\label{fig-vu0}
The computed DR rate is shown for the different initial rovibrational states
of the HeH$^+$ isotopologues, for an initial ground vibrational state $\nu' = 0$.
The computed results are convolved over the electron beam energy distributions
($\Delta E_{||}$ = 0.05 meV and $\Delta E_{\perp}$ = 1.5 meV).
}
\end{figure}

The results in this section have not been averaged over the initial rovibrational population
of the cations. However, the electron beam energy distributions are still averaged over 
($\Delta E_{||}$ = 0.05 meV and $\Delta E_{\perp}$ = 1.5 meV). The beam distribution
widths used here are chosen to be smaller than those in the previous section.
The reason for this change follows from the fact that there are no experimental
data of this type available currently and thus the present data constitute a 
prediction. Ideally, such a prediction should be made for the expected experimental conditions,
such as for the conditions available in the Cryogenic Storage Ring (CSR) at the Max-Planck Institute for Nuclear Physics
\cite{CSR_review_2016}.

Fig.~\ref{fig-vu0} shows the DR rate for the initial vibrational state $\nu' = 0$
and the lowest rotational states. The $\Delta E_{\perp}$ = 1.5 meV is too large
to observe any fine structure close to the first rotational thresholds. What is clear, however, is
that for the ground initial state $\nu'=0, j'=0$, the shape of the DR rate differs
significantly when compared with the Boltzmann-averaged results presented
in Fig.~\ref{fig-isotopes}. For example, if the temperature of the 
cations is in the range of
10--15 K ($\approx$ 1 meV), the population of rotationally excited initial states with
$j' > 0$ can be neglected and the DR rate should be dominated by the black curves.

\subsection{\label{ssec-dr-j1}First rotational and vibrational thresholds}

\begin{figure}[ht]
\begin{center}
\includegraphics[width=0.8\linewidth]{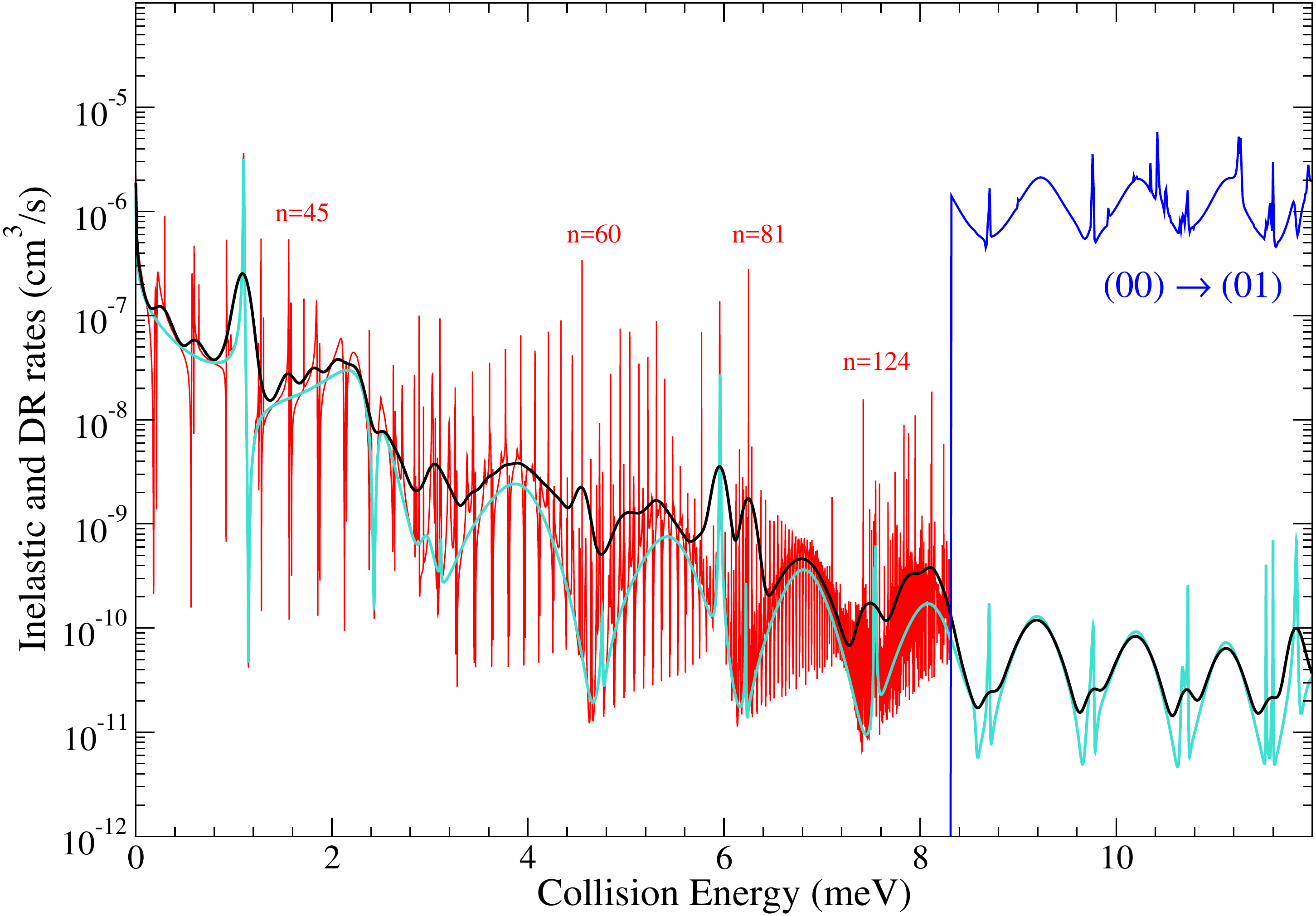}
\end{center}
\caption{\label{fig-j1}
The computed DR rate from the ground ionic rovibrational state is shown in the vicinity of the first rotational
threshold $(01)$. The red line shows the vibrational Feshbach resonances converging
to the first rotational threshold. The turquoise curve is unphysical, but serves a useful purpose in our analysis: it was obtained
by artificially opening the $j = 1$ channel. Note that the red and turquoise curves
coincide above the $j = 1$ threshold. The black curve is a simple one-dimensional
convolution of the red-curve data with a Gaussian function (width of 0.1 meV). The blue
curve displays the rotationally-inelastic rate for the $(00)\rightarrow (01)$ transition.
}
\end{figure}

Fig.~\ref{fig-j1} shows the zoomed-in DR rate from the ground rovibrational state near the
first inelastic threshold characterized by quantum numbers $\nu=0$ and $j=1$.
The turquoise curve displays an unphysical DR rate, corresponding to a process in
which we forbid the scattering electron to be captured in a Rydberg state while the molecule
is rotationally excited to ($\nu,j$) = (0,1).
Such computational experiments can be easily done by opening
artificially the corresponding closed Rydberg channel. The resulting comparison can provide information
about the importance of rotations during the DR process. Fig.~\ref{fig-j1} demonstrates that 
the enhancement
of the DR rate just under the threshold (at $\approx$ 8 meV) can be estimated to
be approximately a factor of 2.
On the other hand, the blue curve showing the inelastic rate for the rotational
transition 0$\rightarrow$1
indicates that rotational excitation is a process about 4 orders of magnitude more probable 
than dissociation. 
\begin{figure}[th]
\begin{center}
\includegraphics[width=0.8\linewidth]{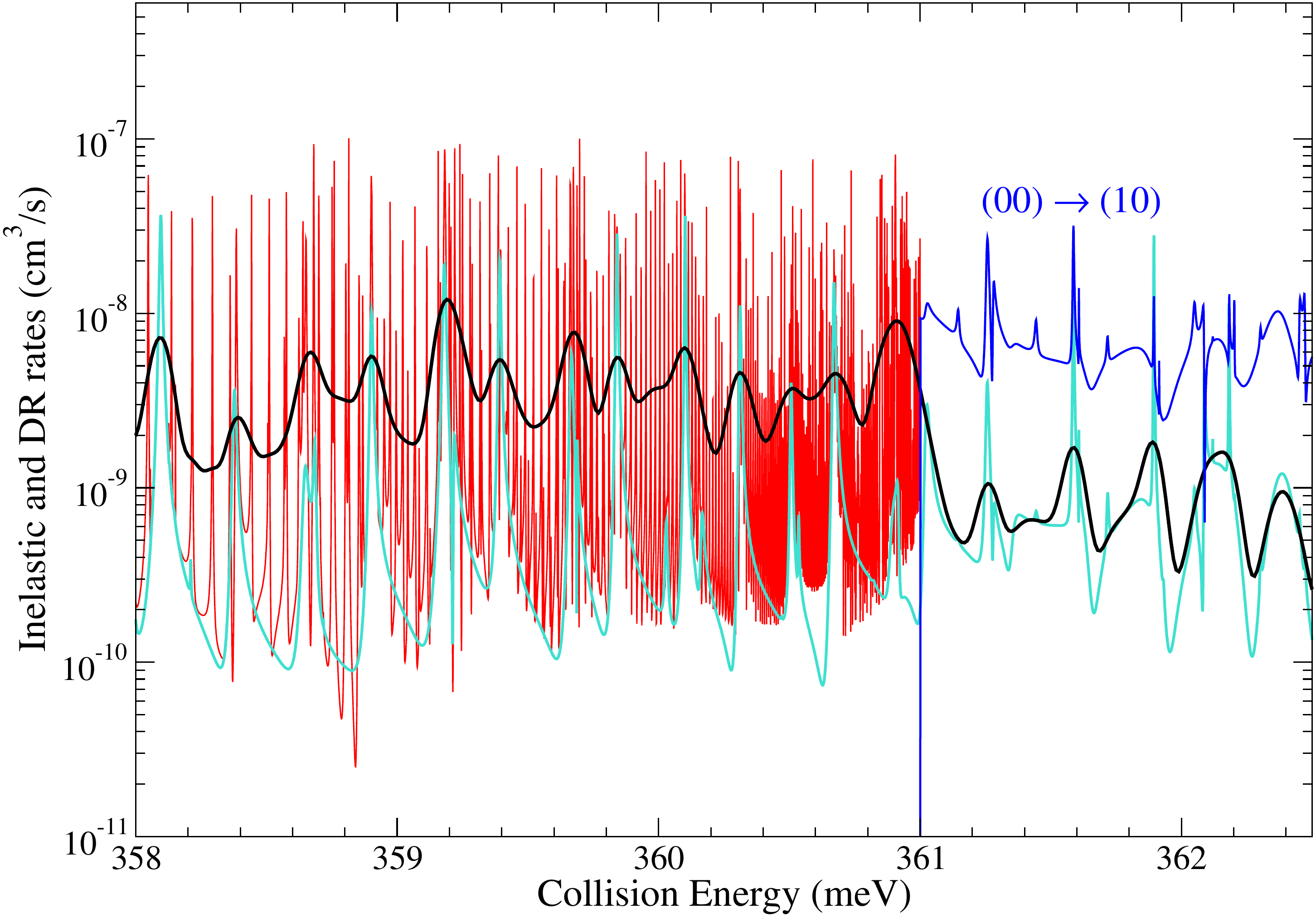}
\end{center}
\caption{\label{fig-v1}
The DR rate from the ground rovibrational state computed in the vicinity of the first vibrational
threshold $(10)$. The coloring scheme is the same as in Fig.~\ref{fig-j1}, except that the
blue line displays here the vibrationally inelastic rate for the
$(00)\rightarrow (10)$ transition.
}
\end{figure}
One can, therefore, conclude that the electrons with these low energies
(say under 8 meV) very efficiently heat the HeH$^+$ cations rotationally by getting trapped
in high Rydberg states (with $n >$ 40). However, these temporary, rotationally excited,
neutral HeH molecules very rarely dissociate, only about 1 case out of 10,000, probably because the
rotational excitation does not excite the molecule along a dissociative coordinate.
This hypothesis can be further examined by a near-threshold inspection, similar to the one shown
in Fig.~\ref{fig-j1}, but this time around the first vibrational threshold.
The corresponding data are displayed in Fig.~\ref{fig-v1}. 
In contrast to the previous
case, closed-channel resonances attached to the first
vibrational threshold enhance the DR rate by about one order of magnitude. This
enhancement is revealed as the difference between the averaged DR rate 
(black curve) and the turquoise curve. The enhancement disappears once the collision
energy is high enough to open the vibrational excitation channel. Fig.~\ref{fig-v1}
also shows that the DR probability just below the threshold transfers
into vibrational excitation probability at energies above the threshold.
Stated in more physical language, this also means that about 90\% of the
temporary, vibrationally excited, neutral high-$n$ HeH molecules that are created by 
below-threshold collisions of electrons with HeH$^+$ will dissociate.

This behavior is similar to what was observed in the case of LiH$^+$
\cite{Curik_Greene_PRL_2007} and it may be a universal property of the 
indirect DR mechanism.

\subsection{\label{ssec-dr-thoughts}More about the DR mechanism}

\begin{figure}[th]
\begin{center}
\includegraphics[width=0.8\linewidth]{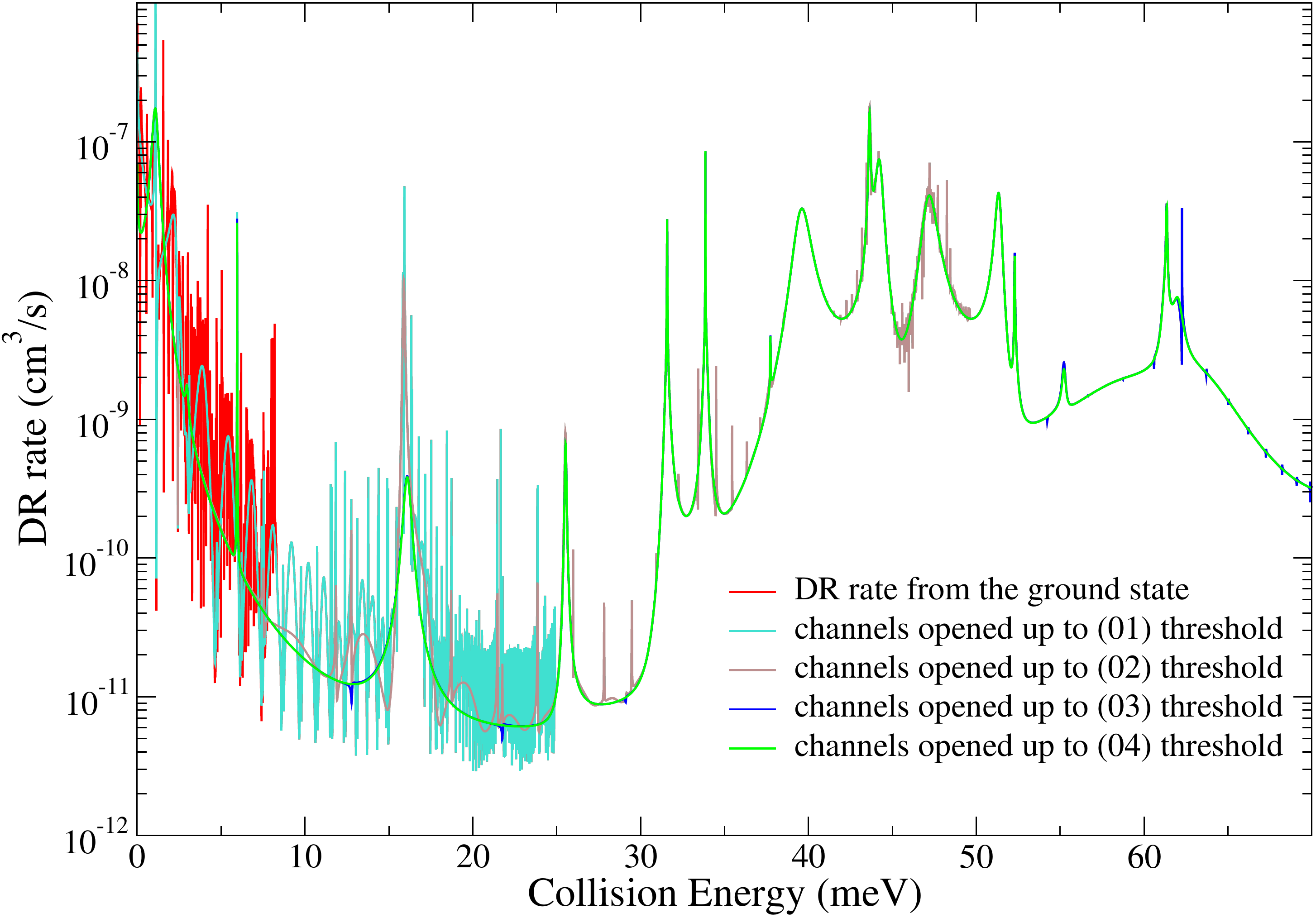}
\end{center}
\caption{\label{fig-win70a}
The computed DR rate from the ground initial state (red curve). The other curves display
calculated unphysical data in which all the closed channels are opened up to
different rotational thresholds ($\nu,j$) (the thresholds are included
in the artificial channel opening procedure).
The colors overlap in the listed order. The red curve is at the bottom while the green curve
is the top color.
}
\end{figure}

\begin{figure}[th]
\begin{center}
\includegraphics[width=0.8\linewidth]{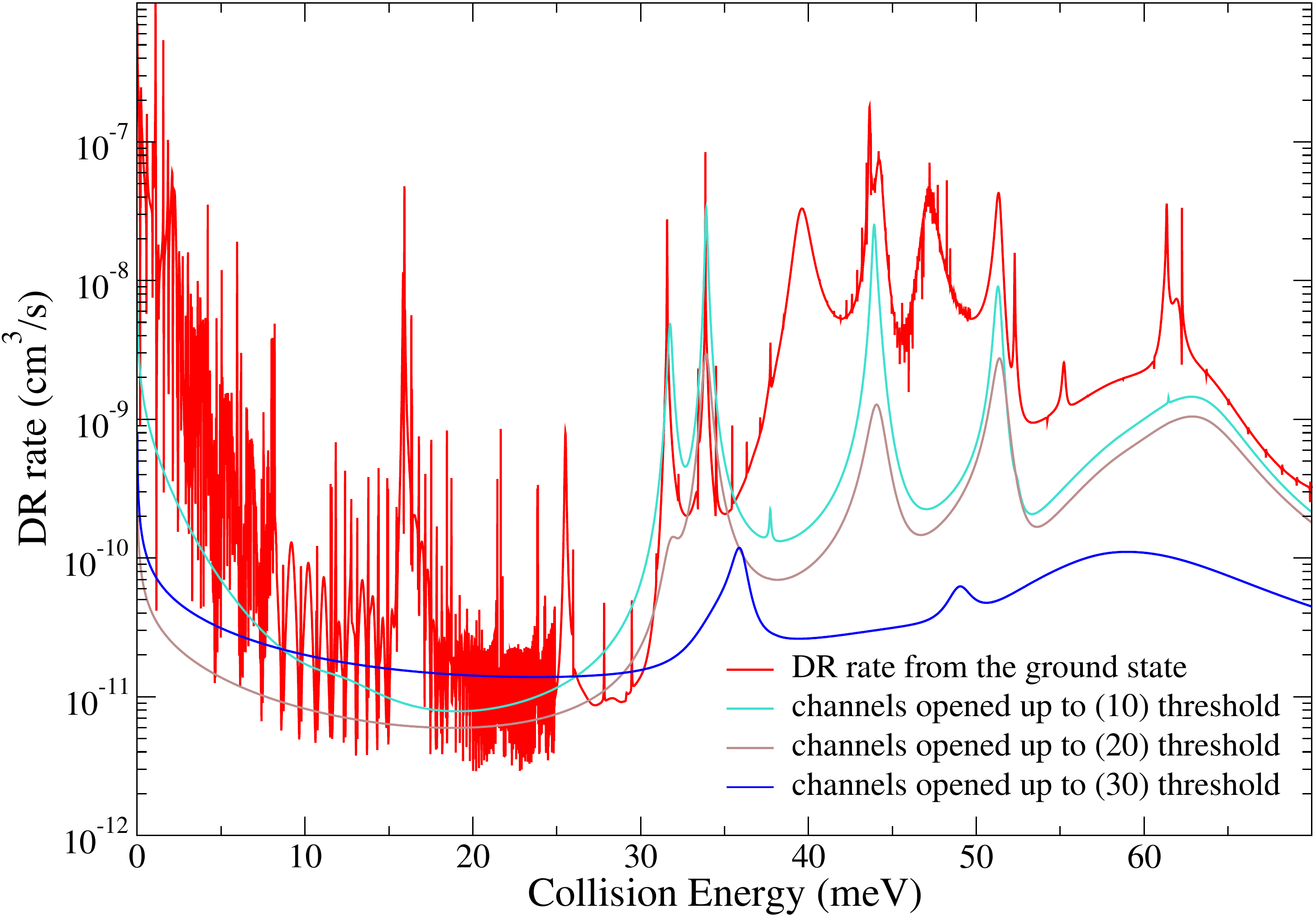}
\end{center}
\caption{\label{fig-win70b}
The computed DR rate from the ground initial state (red curve). Other curves display
calculated unphysical data in which all the closed channels are opened up to
different vibrational thresholds ($\nu,j$) (i.e., the thresholds are included
in the artificial channel-opening procedure).
}
\end{figure}

The preceding paragraphs demonstrate that rotational excitation of the HeH$^+$ during
its collisions with an electron play only a minor role in the DR process for energies near the
first rotational and vibrational thresholds. In light of this, one might still wonder
what is causing the sharp DR enhancements below the first vibrational threshold as 
can be seen in Fig.~\ref{fig-vu0}. As an example we explore the DR rate
for $^4$HeH$^+$ from
the ground rovibrational state (black curve in top left panel in Fig.~\ref{fig-vu0}).
The computed
DR rate, zoomed-in to 0--70 meV, is displayed by the red curve in Fig.~\ref{fig-win70a}.
This figure is just
an energy extension of Fig.~\ref{fig-j1} calculated at a coarser grid.
Fig.~\ref{fig-win70a} also shows results of unphysical calculations in which all
the closed channels are artificially opened up to a certain threshold ($\nu,j$).
Only the first four rotational thresholds are opened but we
did not observe any dominant change in the displayed energy window when the remaining
rotational thresholds (up to (0,12) included in the calculations) were opened.
Fig.~\ref{fig-win70a} suggests that the enhancement of the DR rate around
16 meV and 40--50 meV is not caused by closed-channel resonances attached to
rotationally excited states.

These results motivated a second theoretical experiment in which all the closed channels are
artificially opened up to different vibrational thresholds ($\nu,0$). Corresponding DR rates are
displayed in Fig.~\ref{fig-win70b}. It is quite clear that already opening the first
vibrational threshold (turquoise curve) removes the resonance at 16 meV and
another 2 resonances in the 40--50 meV domain. Artificial opening of the higher
vibrational thresholds removes more of the resonant structure and flattens the
DR rate in the examined energy window.

One can, therefore, conclude that the
enhancement of the DR rate (by 2--3 orders of magnitude) at 40--50 meV is caused
by overlapping of the closed-channel resonances attached to different vibrational
thresholds - a behavior similar to what was previously observed for the LiH system
\cite{Curik_Greene_PRL_2007,Curik_Greene_MP_2007}.

\section{\label{sec-conc}Conclusions}

The calculated data and insights are derived from an {\it ab initio} study describing 
inelastic collisions between low-energy electrons and the HeH$^+$ cation.

One of the goals of the present work has been to address how one’s choice
of the frame of reference affects the computed results. This important
question was previously raised by Takagi and Tashiro 2015
\cite{Takagi_Tashiro_2015}, where the authors observed a difference of about
2 orders of magnitude in the DR rates when using the center-of-mass (CMS) versus
the center-of-charge (CCS) frames of reference.
We have not been able to confirm this conclusion, as we observe differences smaller
than 30\% for collision energies below 100 meV (Fig.~\ref{fig-cmsccs}). Moreover,
the difference in the DR rates becomes negligible for collision energies above
100 meV. The difference is somewhat more pronounced (up to a factor of 2) in the
case of the rotationally inelastic collisions, especially for the $0 \rightarrow 1$
and $0 \rightarrow 3$ transitions (left panel of Fig.~\ref{fig-rvt}). The strong 
$0 \rightarrow 2$ transition, dominated by the frame-invariant quadrupole moment 
of the cation, remains unchanged by the frame of reference change.

Our inelastic rates indicate that rotational excitation is a process almost two orders
of magnitude more probable than vibrational excitation. Several
rotational transitions are compared, after carrying out the thermal-averaging convolution, with the recent
calculations of Hamilton {\it et al.} 2016 (right panel of Fig.~\ref{fig-rvt}).
The agreement with the previous calculations is quite good, with notable differences
for the $0\rightarrow 1$ and $0 \rightarrow 3$ transitions. We assume that these 
differences are caused by the Born-closure technique applied in the previous study.

For the case of dissociative recombination, the computed data reproduce most of
the structures visible in the available experiments for all the HeH$^+$
isotopologues (Fig.~\ref{fig-isotopes}). Furthermore, the absolute experimental 
DR rate values, available for the isotopologues $^4$HeH$^+$ and $^3$HeH$^+$,
agree well with the present calculations.
Our further analysis of the DR mechanism has revealed that while the rotationally
excited neutral high-$n$ HeH molecules form long-lived resonant states, especially when created by below-threshold rotationally inelastic
collisions of electrons with the cations,
these temporary neutral states nevertheless do not contribute much to the dissociation probability.

The situation is very different for the temporary neutral high-$n$ HeH molecules,
created by below-threshold vibrationally inelastic collisions of electrons with the
HeH$^+$ cations. These resonant states lead to dissociation with very high probability
(about 90\%). Since the similar behavior was observed before 
for the LiH$^+$ molecule \cite{Curik_Greene_PRL_2007}, it 
may represent a universal mechanism common for all the indirect DR processes.

Finally, we attribute the strong DR enhancements visible below 100~meV to
the overlap of complex resonance manifolds that couple closed vibrational channels having 
both high and low principal quantum numbers $n$. These overlaps are of the same
type that have been studied in molecular Rydberg spectroscopy in recent decades
\cite{Jungen_Dill_1980,Du_Greene_1986,Greetham_Merkt_2003,Texier_Jungen_2000}.

\begin{acknowledgments}
R\v{C} conducted this work within the COST Action CM1301
(CE\-LI\-NA) and the support of the Czech Ministry of Education (Grant No. LD14088) is
acknowledged.  The work of CHG is supported in part by the U.S. Department of Energy, Office of Science, under Award No. DE-SC0010545.  We thank O. Novotny for his encouragement to carry out this investigation.
\end{acknowledgments}

\bibliographystyle{apsrev}
\bibliography{heh}

\end{document}